\documentclass[aps,pra,amsfonts,twoside,twocolumn,amssymb,superscriptaddress]{revtex4}

\usepackage[latin1]{inputenc}
\usepackage[english]{babel}
\usepackage[draft]{hyperref}
\usepackage{graphicx}
\usepackage{amsmath}
\usepackage{color}
\usepackage{float}

\DeclareMathOperator{\Var}{Var}
\newcommand{\op}[1]{\hat{#1}}
\newcommand{\fiftyfifty}{$50$\,:\,$50$ }

\begin{document}

\title{Strong Einstein-Podolsky-Rosen entanglement from a single squeezed light source}

\author{Tobias Eberle}
\affiliation{Max-Planck-Institut f\"ur Gravitationsphysik
(Albert-Einstein-Institut) and\\ Institut f\"ur Gravitationsphysik
der Leibniz Universit\"at Hannover, Callinstraße 38, 30167 Hannover,
Germany}
\affiliation{Centre for Quantum Engineering and
Space-Time Research - QUEST, Leibniz Universit\"at Hannover,
Welfengarten 1, 30167 Hannover, Germany}
\author{Vitus H\"andchen}
\affiliation{Max-Planck-Institut f\"ur Gravitationsphysik
(Albert-Einstein-Institut) and\\ Institut f\"ur Gravitationsphysik
der Leibniz Universit\"at Hannover, Callinstraße 38, 30167 Hannover,
Germany}
\author{J\"org Duhme}
\affiliation{Centre for Quantum Engineering and
Space-Time Research - QUEST, Leibniz Universit\"at Hannover,
Welfengarten 1, 30167 Hannover, Germany}
\affiliation{Institut f\"ur Theoretische Physik der Leibniz Universit\"at Hannover, Appelstraße 2, 30167 Hannnover}
\author{Torsten Franz}
\author{Reinhard F. Werner}
\affiliation{Institut f\"ur Theoretische Physik der Leibniz Universit\"at Hannover, Appelstraße 2, 30167 Hannnover}
\author{Roman Schnabel}
\email[corresponding author: ]{roman.schnabel@aei.mpg.de}
\affiliation{Max-Planck-Institut f\"ur Gravitationsphysik
(Albert-Einstein-Institut) and\\ Institut f\"ur Gravitationsphysik
der Leibniz Universit\"at Hannover, Callinstraße 38, 30167 Hannover,
Germany}

\begin{abstract}
Einstein-Podolsky-Rosen (EPR) entanglement {is} a criterion that is more demanding than just certifying entanglement. We theoretically  and experimentally analyze the low resource generation of bi-partite continuous variable entanglement, as realized by mixing a squeezed mode with a vacuum mode at a balanced beam splitter, i.e. the generation of so-called vacuum-class entanglement. We find that in order to observe {EPR} entanglement the total optical loss must be smaller than $33.3$\,$\%$. However, arbitrary strong EPR entanglement is generally possible with this scheme. We realize continuous wave squeezed light at $1550$~nm with up to $9.9$~dB of non-classical noise reduction, which is the highest value at a telecom wavelength so far. Using two phase controlled balanced homodyne detectors we observe an EPR co-variance product of  $0.502\pm0.006 < 1$, where $1$ {is the critical value}. We discuss the feasibility of strong Gaussian entanglement and its application for quantum key distribution in a short-distance fiber network.
\end{abstract}

\maketitle

\section{Introduction}
In this paper we explore entanglement in the quadrature measurements on a pair of laser beams. This is a concrete realization of the system considered in 1935 by Einstein, Podolsky and Rosen (EPR) \cite{EPR1935}: a two-mode system with a canonical pair of continuous variables measured on each side, in an overall Gaussian state. It might appear that this system was ill-chosen for the investigation of non-classical features of quantum mechanics. After all, in this situation the Wigner function immediately provides a classical probabilistic model for all measurements involved, and hence no Bell inequality can be violated in this setup. Nevertheless, EPR established some non-classical features in such a state, for which a quantitative criterion was proposed by M. Reid \cite{Reid1989}. Her criterion,``EPR entanglement'' captures a distinction, which is meaningful also in a broader context and has been called ``steering'' \cite{Wiseman07}, after another term used by Schr\"odinger. For our paper it is important that this criterion is more demanding than just establishing entanglement \cite{Wiseman09}. Moreover, it is applicable without an assumption about the Gaussian nature of the state.

\section{Background}
Following Reid, we call a state EPR entangled, if
\begin{equation}
\label{eq:EPR}
\Var_{A|B}(\op{X}_A, \op{X}_B) \cdot \Var_{A|B}(\op{P}_A, \op{P}_B) < 1\ ,
\end{equation}
where $\op{X}$ and $\op{P}$ are the amplitude quadrature (position) and phase quadrature (momentum) operators, respectively and $\Var_{A|B}$ denote conditional variances. These are defined as $\Var_{A|B}(\op{X}_A, \op{X}_B) = \min_g\Var(\op{X}_A-g\op{X}_B)$, where the parameter $g$ is varied to minimize the variance of the discrepancy of the measured and the inferred outcomes of the measurement.

EPR entangled states were e.g.\ generated in~\cite{Ou1992}\nocite{Zhang2000,Schori2002,Laurat2005}\,-\cite{Keller2008} by type II parametric down-conversion, in~\cite{Bowen2003,Takei2006} by type I parametric down-conversion, and in ~\cite{Silberhorn2001} by the optical Kerr effect. In all these experiments the nonclassical resource of the EPR entanglement generation could be described by the interference of two squeezed modes. In the case of type I parametric down-conversion, two squeezed modes were generated in two independent nonlinear cavities and subsequently overlapped on a balanced beam splitter.

In \cite{AFLP66,LBr00} it was theoretically shown that an entangled state is generated from a pure single squeezed mode for any nonzero squeezing. EPR entanglement from a single squeezed mode was theoretically analyzed in \cite{BLR03}. However, experimental imperfections such as optical loss were not considered in these works and so far EPR entanglement from a single squeezed mode were not experimentally demonstrated.

Here we report on the generation of EPR entanglement from the interference of a squeezed mode with a vacuum mode, which we call v-class entanglement \cite{DiGuglielmo2007}. For pure v-class entanglement the variance product (\ref{eq:EPR}) is given by 
\begin{equation}
\label{eq:EPRPure}
\frac{4}{2+\Var_{\rm{asqz}}+\Var_{\rm{sqz}}} < 1\ 
\end{equation}
where $\Var_{\rm{(a)sqz}}$ describes the variance of the (anti)squeezed quadrature of the pure state normalized to the vacuum noise variance. With sufficiently high squeezing this can become arbitrarily small, while the entanglement, as measured by the entropy of a subsystem, diverges. Since just a single squeezed mode is required as the input beam of a balanced beam splitter, the experimental setup is less involved compared to previous experiments. We observe a significant EPR entanglement yielding a value of $0.502 \pm 0.006$ according to Eq.~(\ref{eq:EPR}) and discuss the influence of optical loss on our setting.

For the symmetric situation, i.e.\ with two equally squeezed states as inputs to a balanced beam splitter (so-called \emph{s-class} entanglement \cite{DiGuglielmo2007}), it is known that EPR entanglement can only be observed if the total optical loss on the state is smaller than $50\%$, as experimentally demonstrated in Ref.~\cite{Bowen2003}. In the case of v-class entanglement and under the assumption of symmetric optical loss, EPR entanglement is observed, if
\begin{equation}
  \left(1-\Var_{\rm{sqz}}\right)^2(1-\mu)\left(\frac{1}{3}-\mu\right) > 0\ .
\end{equation}
Here $\mu$ describes the total optical loss of the initially pure state. This condition shows that the restriction on the optical loss for v-class entanglement is more severe than for s-class entanglement. The loss on v-class entangled states has to be smaller than $33.3\%$ in order to be able to observe EPR entanglement.
In contrast `inseparability' is present in this scheme for any loss $< 100\%$.

\section{Demonstration of EPR Entanglement}

\begin{figure}[ht]
  \includegraphics[width=8.2cm]{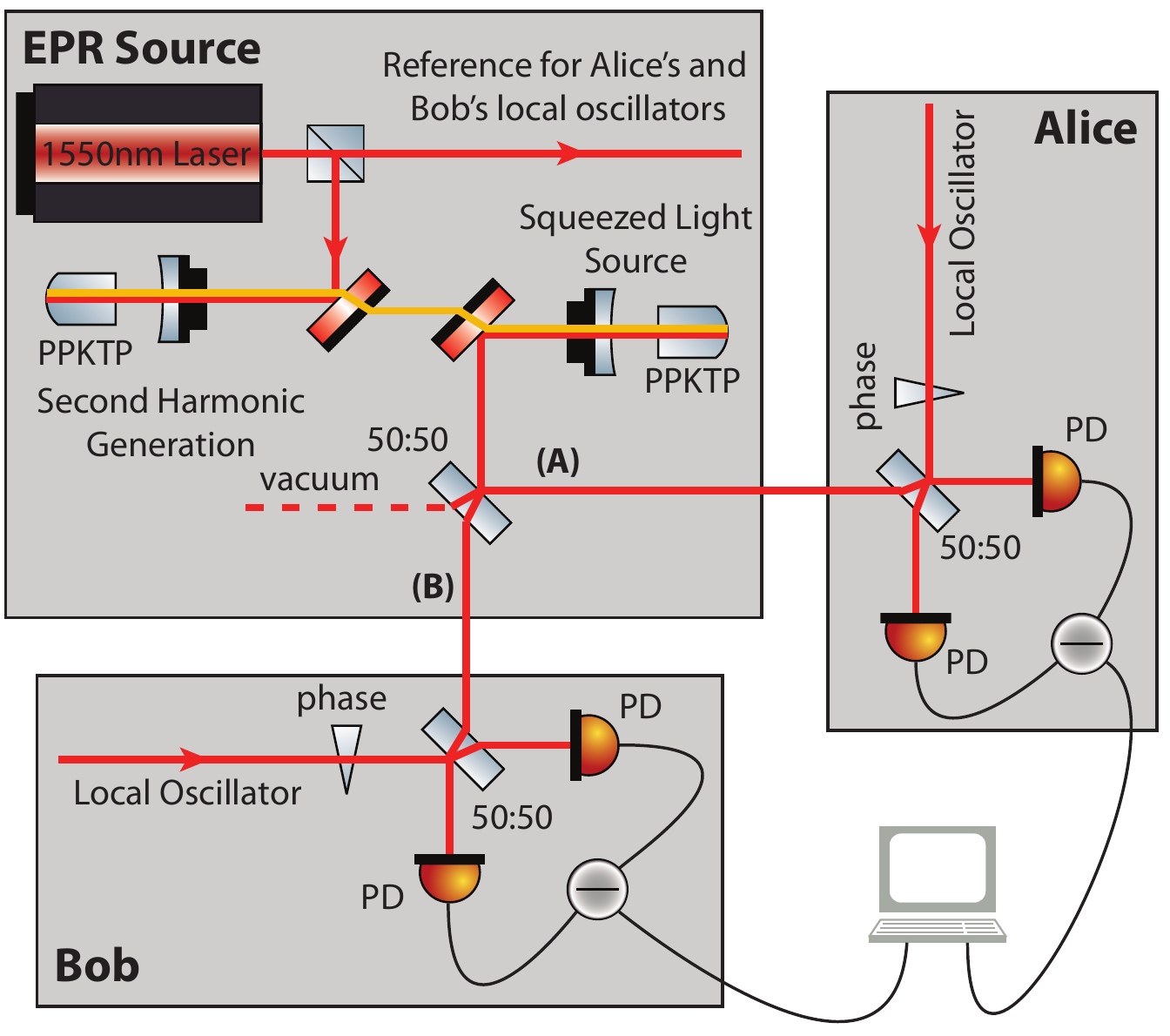}
  \caption{(Color online) Schematic of the experiment. A continuous-wave laser beam at $1550$\,nm (red) was frequency doubled (yellow) and used to produce squeezed light at the fundamental wavelength via type I parametric down-conversion. A balanced beam splitter subsequently mixed the squeezed light with a vacuum mode. Amplitude and phase quadrature amplitude measurements on the output beams by means of balanced homodyne detectors proved EPR entanglement amongst them. PD: photo diode.
  }
  \label{fig:experiment}
\end{figure}

Figure~\ref{fig:experiment} shows a schematic of the experimental setup. The quadrature amplitudes of both output beams of the \fiftyfifty entanglement beam splitter were detected by means of  balanced homodyne detectors (Alice and Bob) consisting of two high quantum efficiency ($> 95\%$) InGaAs photo diodes (PD) each. The phases of the local oscillators of the homodyne detectors were locked either to the amplitude or to the phase quadrature of the field. The outcomes were recorded simultaneously by a data acquisition system at a sample rate of $500$\,kHz. For this purpose the signals were mixed down at a frequency of $5$\,MHz using a double balanced mixer and a lowpass filter at $35$\,kHz which served also as anti-aliasing filter for the $16$\,bit data acquisition system. For each measurement $5 \times 10^6$ data points were sampled.

The squeezed light was produced by type I parametric down-conversion in a periodically poled potassium titanyl phosphate (PPKTP) crystal. One end-face of the $9.3$\,mm long crystal was curved with a radius of curvature of $12$\,mm forming a half-monolithic cavity together with an external coupling mirror with a radius of curvature of $25$\,mm. The curved end-face of the crystal had a high-reflectivity coating for both the fundamental at $1550$\,nm and the pump at $775$\,nm, whereas the other end-face was coated anti-reflective. The coupling mirror had a reflectivity of $R = 90$\,\% for the fundamental and $R = 20$\,\% for the pump and could be actuated by means of a piezo-electric transducer (PZT) to keep the cavity on resonance. The crystal was temperature controlled to $35^\circ$\,C to achieve phase-matching.

The main laser source of the experiment was an $1$\,W fiber laser at $1550$\,nm. Its beam served as pump for a second harmonic generation, made of PPKTP as well, to generate the $775$\,nm pump beam required for the squeezed light source. A small fraction of the main laser beam was filtered by a mode-cleaning ring cavity with a finesse of about $300$ and served as the homodyne detector local oscillators.


\begin{figure}[ht]
  \includegraphics[width=8.2cm]{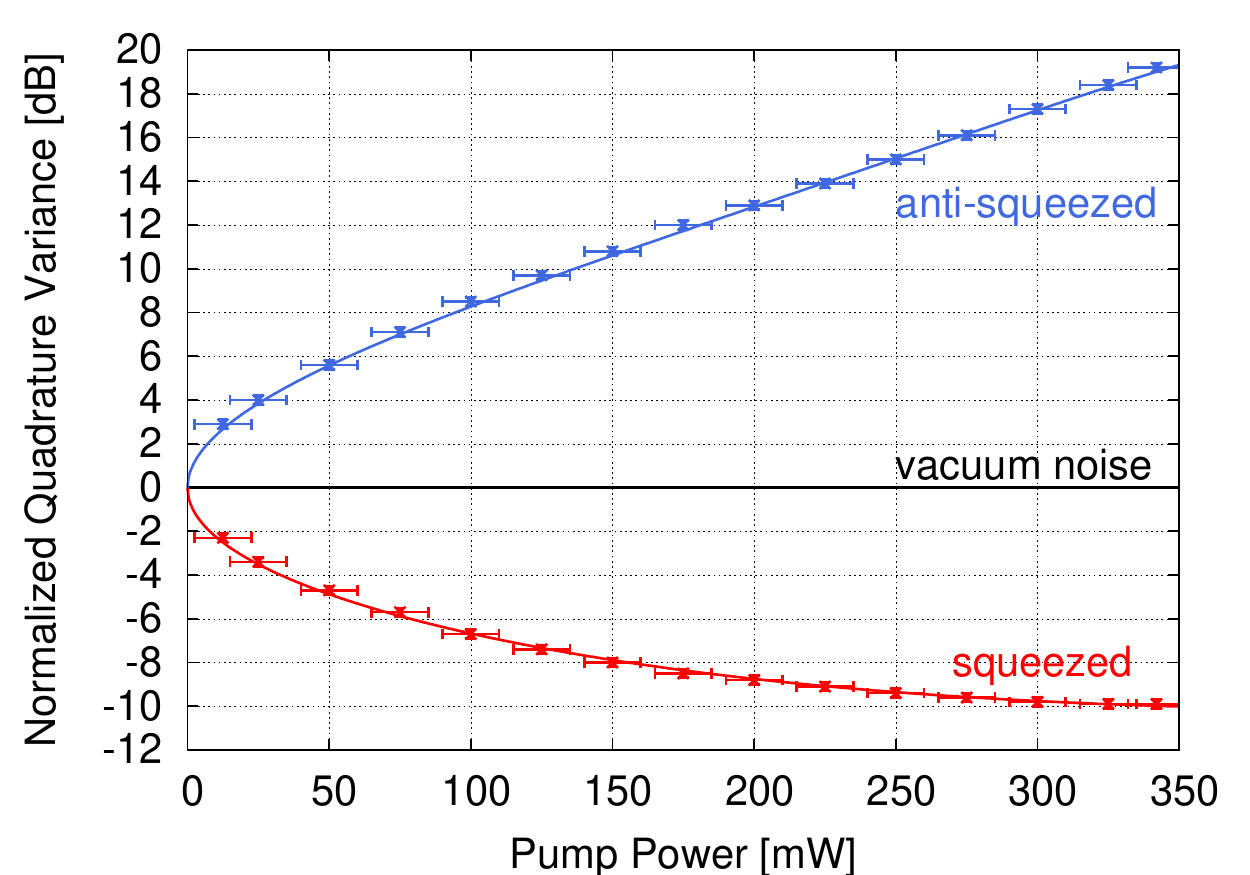}
  \caption{(Color online) Characterization of our squeezed light laser at a sideband frequency of $5$~MHz. Shown are the vacuum noise normalized squeezed and anti-squeezed quadrature variances for several light powers of the pump field. The variances were measured by a balanced homodyne detector and averaged ten times. The detector's dark noise was at $-22$~dB and therefore irrelevant, and not subtracted from the data. The solid lines show a model for our data according to Eq.\,(\ref{eq:sqz_var}).
  }
  \label{fig:sqz}
\end{figure}

Figure~\ref{fig:sqz} presents a characterization of our squeezed light source. For this measurement the entanglement beam splitter was removed and the squeezed field directly characterized by balanced homodyne detection. The graph shows the variance of the squeezed and anti-squeezed quadrature relative to the vacuum noise versus the pump power for the nonlinear process, given in decibel. For a pump power of $325$\,mW we observed $9.9$\,dB of squeezing and $18.4$\,dB of anti-squeezing. To the best of our knowledge this is the highest squeezing value ever observed at a wavelength of $1550$\,nm~\cite{Mehmet2009,Mehmet2010}. Squeezing levels larger than $10$\,dB were observed in Refs.~\cite{Vahlbruch2008,Mehmet2010a,Eberle2010} at $1064$\,nm. The solid lines in the figure represent a theoretical model. The squeezed (sqz) and anti-squeezed (asqz) quadrature variances of the field can be described as a function of pump power $P$ by~\cite{Takeno2007}
\begin{equation}
  \label{eq:sqz_var}
  \Var_{\rm{sqz,asqz}} = 1\pm \eta\gamma\frac{4\sqrt{P/P_\text{th}}}{(1\mp\sqrt{P/P_\text{th}})^2 + 4 K(f)^2}\ ,
\end{equation}
where $\eta$ is the detection efficiency and $\gamma$ the escape efficiency of the nonlinear cavity. $P_{\rm{th}}$ is the threshold power and $K(f) = 2\pi f/\kappa$ the ratio between Fourier frequency $f = 5$\,MHz and the cavity decay rate $\kappa = (T+L)c/l$ with the output coupler transmission $T$, the intra-cavity loss $L$, the speed of light in vacuum $c$ and the cavity round trip length $l = 79.8$~mm. The model fits best with a total optical loss of $1-\eta\gamma = 0.09$, a threshold power of $P_{\rm{th}}=445$\,mW and $T+L = 0.105$.

\begin{figure}[ht]
  \includegraphics[width=8.2cm]{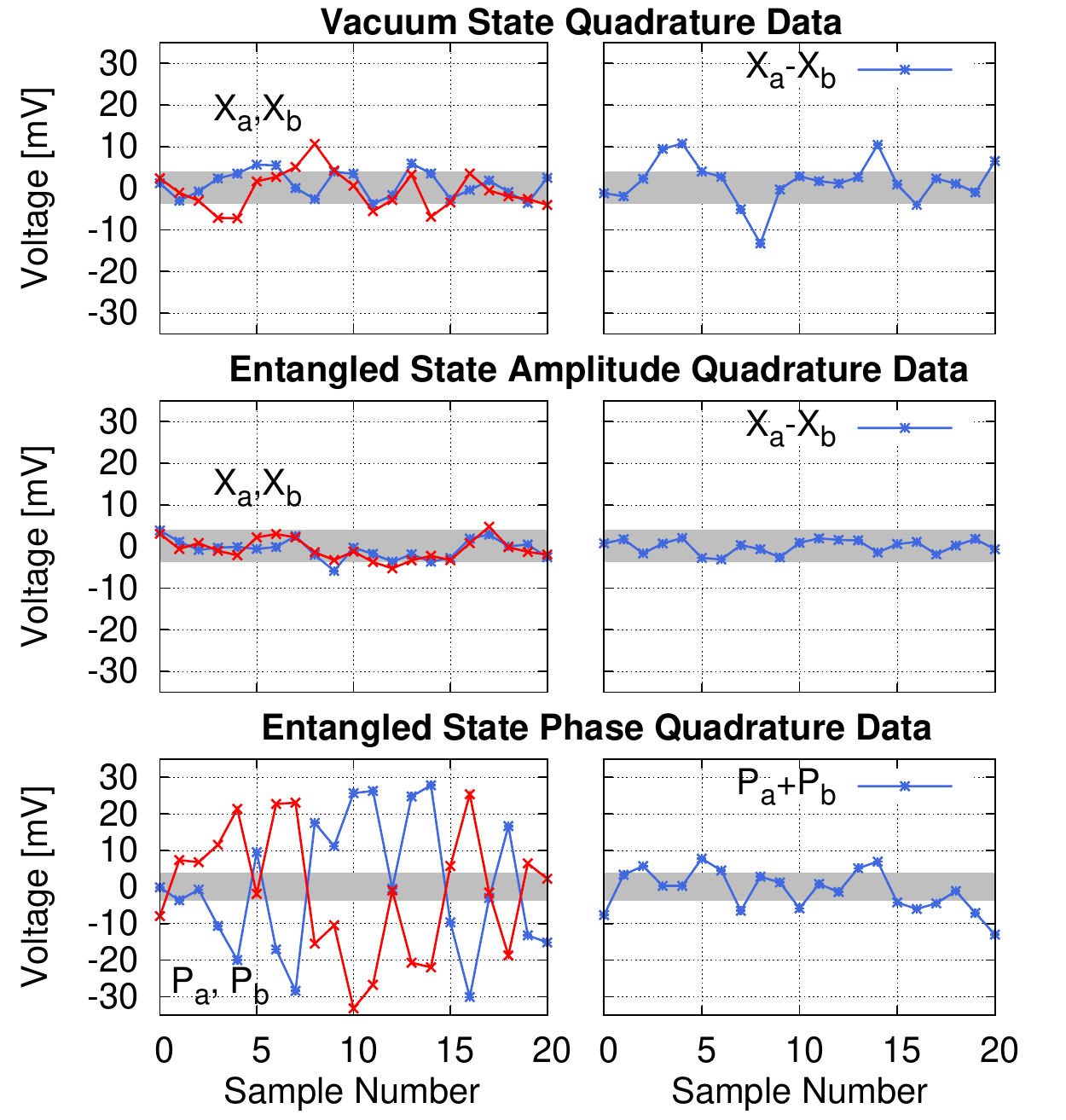}
  \caption{(Color online) Typically time series of measured quadrature amplitude data. The first row shows the voltages measured by Alice's and Bob's homodyne detector for a vacuum state input, whereas the lower two are for a v-class entangled state and its amplitude ($X$) and phase quadrature ($P$), respectively. The grey areas indicate the standard deviation of the vacuum noise for a single mode.
  }
  \label{fig:timeseries}
\end{figure}

The EPR entanglement of the two light fields A and B produced by a single squeezed field overlapped with a vacuum mode was characterized by two homodyne detectors, i.e. the two observers Alice and Bob. They simultaneously took either amplitude quadrature or phase quadrature data. Typical data is shown in Fig.~\ref{fig:timeseries}. For the measurement in the first row the output of the squeezed light source was blocked and only vacuum noise were present at both receivers. Hence the difference (and the sum, not shown) shows a standard deviation that is a factor of $\sqrt{2}$ above the vacuum noise. The lower two rows show the measurements on a v-class entangled state. The amplitude quadrature data as well as the phase quadrature data taken at Alice's and Bob's sites show correlations. It is clearly visible that the fluctuations in the right panels are less than those in the left panels.

\begin{figure}
  \includegraphics[width=8.2cm]{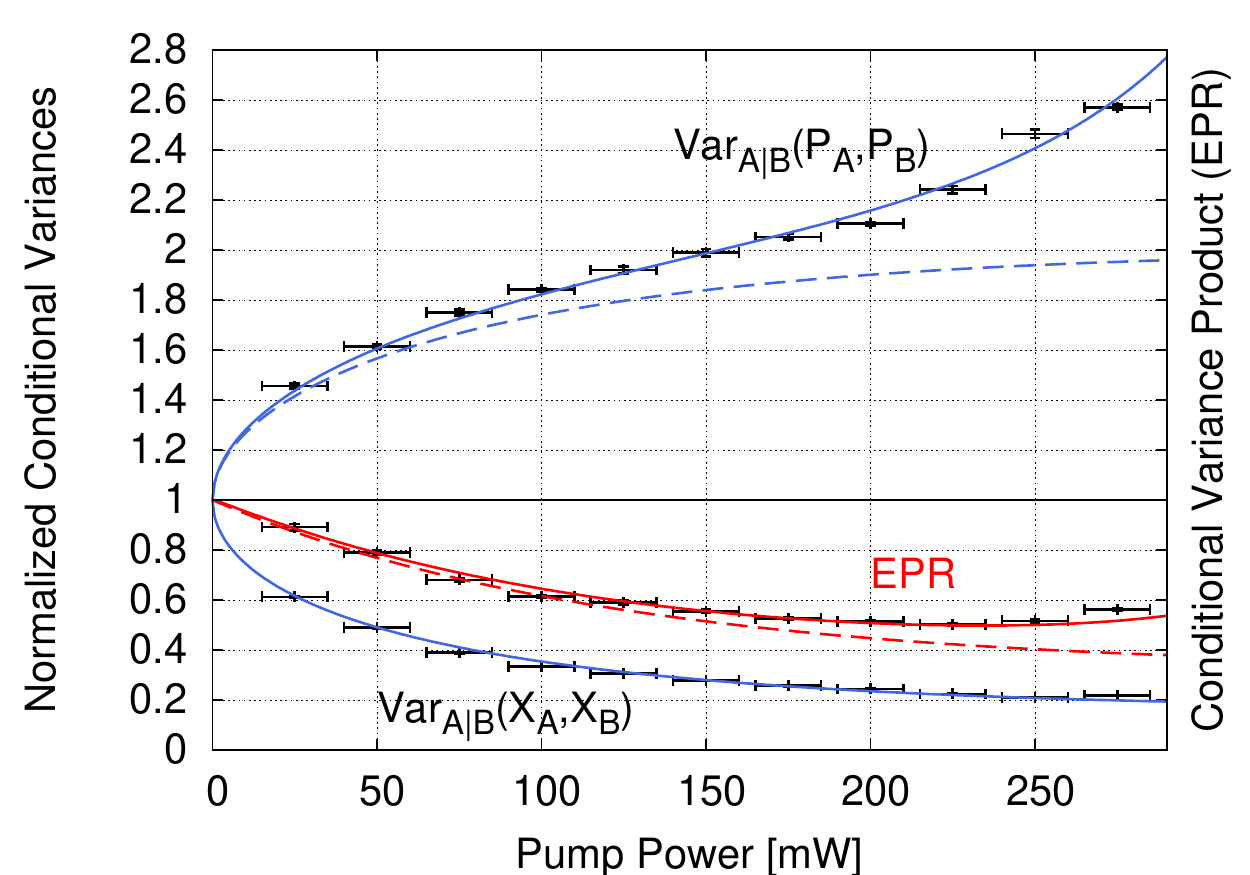}
  \caption{(Color online) Conditional variances for amplitude and phase quadratures and the EPR criterion calculated from the measured data. The dashed lines represent the theoretical model given by equation (\ref{eq:sqz_var}), the solid lines represent the model with additional excess noise. Both models coincide for the conditional variance of the amplitude quadrature.
  }
  \label{fig:epr}
\end{figure}

Figure~\ref{fig:epr} shows the conditional variances and its product for the EPR criterion according to Eq.\,(\ref{eq:EPR}). In general, the roles of Alice and Bob are not interconvertible, i.e., EPR entanglement is not a symmetric quantity. Experimentally we found only minor differences in the phase quadrature, that probably originated in asymmetric optical loss. The differences for the EPR entanglement and the $X$ quadrature measurements would not be visible in the figure, so only one direction is plotted. Our best value for the EPR criterion is $0.502\pm 0.006$ and was achieved for a pump power of $225$~mW. Here we found an entanglement of formation of 1.16~\cite{Rigolin2004}. Given the fact that we used just a single squeezed mode, our result is quite remarkable. To the best of our knowledge only one work reported a stronger EPR entanglement~\cite{Laurat2005}, however, based on type II parametric down conversion and therefore to an equivalent of two squeezed input modes.
The dashed lines in Fig.~\ref{fig:epr} represent the predictions from Eq.\,(\ref{eq:sqz_var}). We observe a considerable difference to the experimental data at higher pump powers. Comparing the data with different noise simulations we could identify excess noise in the (anti-squeezed) phase quadrature as main noise source, shown as solid line in Fig.~\ref{fig:epr}. This excess noise probably arose due to nonlinear response of the detectors at high pump powers. Effects of phase noise were only observed marginally and therefore omitted.

\section{Discussion and Outlook}

The high strength of the entanglement generated from just a single squeezed field at 1550\,nm is an encouraging result. By replacing the vacuum mode in our setup with a second squeezed mode the entanglement strength will further increase considerably. Without additional loss, this would lead to an EPR value of around $ \Var_{A|B}^2=(0.2)^2=0.04 $. 

In the present work we have shown that balanced homodyne detectors can be locked to a certain quadrature phase precisely enough to measure a conditional variance of just 0.2, see lower trace in Fig.~\ref{fig:epr}. Furthermore, we have also shown that the spatial mode of our entangled field can be aligned to spatially filtered local oscillators in the fundamental Gaussian TEM00 mode with a visibility of  $99.5$\,$\%$. From this result we infer that a similar high visibility will be achieved between two squeezed modes as required to generate the envisaged extremely low conditional variance products. In a previous experiment a continuous-wave squeezed field at 1550\,nm was coupled into an optical fiber, transmitted and finally coupled out again in order to measure its nonclassical properties~\cite{Mehmet2010}. This shows that one is able to overcome coupling losses and so fiber-based distribution of squeezing is feasible for distances of several kilometers. Furthermore, the result can directly be transferred to the distribution of Gaussian quadrature entanglement as analyzed here. A possible application is Gaussian quantum key distribution (see e.g. \cite{Scarani09} and references therein).

Simulations show that by replacing the vacuum input with a second squeezed state, key rates around one bit per time-bin will be possible. Since squeezed fields with bandwidths above 100\,MHz have already been demonstrated~\cite{Mehmet2010a} quantum key distribution with rates in the 100 Mbit$/$Sec.-regime should be possible with two-mode squeezed states.

\begin{acknowledgments}
This research was supported by the FP\,7 project Q-ESSENCE (Grant agreement number 248095). TE and VH thank the IMPRS on Gravitational Wave Astronomy for support. TF acknowledges support from DFG under grant WE-1240/12-1.
\end{acknowledgments}


\begin{thebibliography}{99}
\bibitem{EPR1935} A.\,Einstein, B.\,Podolsky, N.\,Rosen, Phys.\,Rev. {\bf 47}, 777 (1935)
\bibitem{Reid1989} M.\,D.\,Reid, Phys.\,Rev.\,A {\bf 40}, 913 (1989)
\bibitem{Wiseman07} H.\,M.\,Wiseman, S.\,J.\,Jones, and A.\,C.\,Doherty, Phys.\,Rev.\,Lett. \textbf{98}, 140402 (2007)
\bibitem{Wiseman09} E.\,G.\,Cavalcanti,  S.\,J.\,Jones, H.\,M.\,Wiseman, M.\,D.\,Reid, Phys.\,Rev.\,A {\bf 80}, 032112 (2009)
\bibitem{Ou1992} Z.\,Y.\,Ou, S.\,F.\,Pereira, H.\,J.\,Kimble, K.\,C.\,Peng, Phys.\,Rev.\,Lett. {\bf 68}, 3663 (1992)
\bibitem{Zhang2000} Y.\,Zhang, H.\,Wang, X.\,Li, J.\,Jing, C.\,Xie, K.\,Peng, Phys.\,Rev.\,A {\bf 62}, 023813 (2000)
\bibitem{Schori2002} C.\,Schori, J.\,L.\,Sorensen, E.\,S.\,Polzik, Phys.\,Rev.\,A {\bf 66}, 033802 (2002)
\bibitem{Laurat2005} J.\,Laurat, T.\,Coudreau, G.\,Keller, N.\,Treps, C.\,Fabre, Phys.\,Rev.\,A {\bf 71}, 022313 (2005)
\bibitem{Keller2008} G.\,Keller, V.\,D'Auria, N.\,Treps, T.\,Coudreau, J.\,Laurat, C.\,Fabre, Opt.\,Express {\bf 16}, 9351 (2008)
\bibitem{Bowen2003} W.\,P.\,Bowen, R.\,Schnabel, P.\,K.\,Lam, T.\,C.\,Ralph, Phys.\,Rev.\,Lett. {\bf 90}, 043601 (2003)
\bibitem{Takei2006} N.\,Takei, N.\,Lee, D.\,Moriyama, J.\,S.\,Neergaard-Nielsen, A.\,Furusawa, Phys.\,Rev.\,A {\bf 74}, 060101(R) (2006)
\bibitem{Silberhorn2001} C.\,Silberhorn, P.\,K.\, Lam, O.\,Weiß, F.\,K{\"o}nig, N.\,Korolkova, G.\,Leuchs, Phys.\,Rev.\,Lett. {\bf 86}, 4267 (2001)
\bibitem{AFLP66} Y.\,Aharonov, D.\,Falkoff, E.\,Lerner, and  H.\,Pendleton, Ann. Phys.  {\bf 39}, 498 (1966)
\bibitem{LBr00} P.\,van Loock and S.\,L.\,Braunstein, Phys.\,Rev.\,Lett. {\bf 84}, 3482 (2000)
\bibitem{BLR03} W.\,P.\,Bowen, P.\,K.\,Lam, and T.\,C.\,Ralph, J.\,Mod.\,Opt. {\bf 50}, 801 (2003)
\bibitem{DiGuglielmo2007} J.\,DiGuglielmo, B.\,Hage, A.\,Franzen, J.\,Fiur\'{a}\v{s}ek, R.\,Schnabel, Phys.\,Rev.\,A {\bf 76}, 012323 (2007)
\bibitem{Mehmet2009} M.\,Mehmet, S.\,Steinlechner, T.\,Eberle, H.\,Vahlbruch, A.\,Th\"uring, K.\,Danzmann, R.\,Schnabel, Opt.\,Lett. {\bf 34}, 1060 (2009)
\bibitem{Mehmet2010} M.\,Mehmet, T.\,Eberle, S.\,Steinlechner, H.\,Vahlbruch, R.\,Schnabel, Opt.\,Lett. {\bf 35}, 1665 (2010)
\bibitem{Vahlbruch2008} H.\,Vahlbruch, M.\,Mehmet, S.\,Chelkowski, B.\,Hage, A.\,Franzen, N.\,Lastzka, S.\,Go{\ss}ler, K.\,Danzmann, R.\,Schnabel, Phys.\,Rev.\,Lett. {\bf 100}, 033602 (2008)
\bibitem{Mehmet2010a} M.\,Mehmet, H.\,Vahlbruch, N.\,Lastzka, K.\,Danzmann, R.\,Schnabel, Phys.\,Rev.\,A {\bf 81}, 013814 (2010)
\bibitem{Eberle2010} T.\,Eberle, S.\,Steinlechner, J.\,Bauchrowitz, V.\,H\"andchen, H.\,Vahlbruch, M.\,Mehmet, H.\,M\"uller-Ebhardt, R.\,Schnabel, Phys.\,Rev.\,Lett.\, {\bf 104}, 251102 (2010)
\bibitem{Takeno2007} Y.\,Takeno, M.\,Yukawa, H.\,Yonezawa, A.\,Furusawa, Opt.\,Express {\bf 15}, 4321 (2007)
\bibitem{Rigolin2004} G.\,Rigolin, C.\,O.\,Escobar, Phys.\,Rev.\,A {\bf 69}, 012307 (2004)
\bibitem{Scarani09} V.\,Scarani, H.\,Bechmann-Pasquinucci, N.\,J.\,Cerf, M.\,Dusek, N.\,L\"utkenhaus, M.\,Peev, Rev.\,Mod.\,Phys. {\bf 81}, 1301 (2009)
\end{thebibliography}
\end{document}